\documentclass[12pt]{article}

\usepackage{epsfig}
\usepackage{subfigure}
\usepackage{german}
\newcommand{\zed} {\mathrm{Z}}

\newcommand{\epem}  {\mathrm{e}^+\mathrm{e}^-}
\newcommand{\qqbar} {\mathrm{q}\bar{\mathrm{q}}}
\newcommand{\uubar} {\mathrm{u}\bar{\mathrm{u}}}
\newcommand{\ddbar} {\mathrm{d}\bar{\mathrm{d}}}
\newcommand{\ssbar} {\mathrm{s}\bar{\mathrm{s}}}
\newcommand{\bbbar} {\mathrm{b}\bar{\mathrm{b}}}

\newcommand{\unit}[1]{\,{\mathrm{#1}}}

\def \MC {Monte Carlo }

\def \pip{ \pi^+ }
\def \pim{ \pi^- }

\def \piz{ \pi^0 }

\def \mm{ \mu^{+} \mu^{-} }

\begin{document}

\begin{titlepage}
\begin{center}
EUROPEAN ORGANIZATION FOR NUCLEAR RESEARCH (CERN)
\end{center}
\begin{flushright}

\bigskip
\bigskip
\bigskip

CERN-EP/2001-090\\
December 11, 2001\\
\end{flushright}

\bigskip
\bigskip
\bigskip
\bigskip
\bigskip

\begin{center}
\setlength{\baselineskip}{0.8cm}
{\LARGE \bf Inclusive Production of the \boldmath $\omega$ and} \\
{\LARGE \bf \boldmath $\eta$ Mesons in Z Decays,} \\
{\LARGE \bf and the Muonic Branching Ratio of the \boldmath $\omega$}
\end{center}

\bigskip
\bigskip
\bigskip

\begin{center}
{\large The ALEPH Collaboration$^{*)}$}
\end{center}

\bigskip
\bigskip
\bigskip


\begin{abstract}
\noindent
The inclusive production of the $\omega(782)$ vector meson
in hadronic Z decays is measured and compared to model predictions.
The analysis is based on 4 million hadronic Z decays recorded by the
ALEPH detector between 1991 and 1995.
The production rate for $x_p = p_{\mathrm {meson}}/p_{\mathrm {beam}} 
> 0.05$ is measured in the $\omega \rightarrow \pi^{+} \pi^{-} \pi^{0}$ 
decay mode and found to be
$0.585 \pm 0.019_{\mathrm {stat}} \pm 0.033_{\mathrm {sys}}$ per event.
Inclusive $\eta$ meson production is also measured in the same decay channel 
for $x_p > 0.10$, obtaining 
$0.355 \pm 0.011_{\mathrm {stat}} \pm 0.024_{\mathrm {sys}}$ per event.
The branching ratio for $\omega \rightarrow \mm$ is
investigated. A total of $18.1 \pm 5.9$ events are observed, from which 
the muonic branching ratio is measured for the first time to be 
BR$(\omega \rightarrow \mu^+ \mu^-) = 
(9.0 \pm 2.9_{\mathrm {stat}} \pm 1.1_{\mathrm {sys}}) \times 10^{-5}$.

\bigskip
\bigskip
\bigskip

\begin{center}
{\it To be submitted to Phys.\ Lett.\ B}
\end{center}

\bigskip

\end{abstract}

\vfill

~~~~~$^*)$ See next pages for the list of authors
\end{titlepage}


\pagestyle{empty}
\newpage
\small
%
\newlength{\saveparskip}
\newlength{\savetextheight}
\newlength{\savetopmargin}
\newlength{\savetextwidth}
\newlength{\saveoddsidemargin}
\newlength{\savetopsep}
\setlength{\saveparskip}{\parskip}
\setlength{\savetextheight}{\textheight}
\setlength{\savetopmargin}{\topmargin}
\setlength{\savetextwidth}{\textwidth}
\setlength{\saveoddsidemargin}{\oddsidemargin}
\setlength{\savetopsep}{\topsep}
%
%
\setlength{\parskip}{0.0cm}
\setlength{\textheight}{25.0cm}
\setlength{\topmargin}{-1.5cm}
\setlength{\textwidth}{16 cm}
\setlength{\oddsidemargin}{-0.0cm}
\setlength{\topsep}{1mm}
\pretolerance=10000
\centerline{\large\bf The ALEPH Collaboration}
\footnotesize
\vspace{0.5cm}
{\raggedbottom
\begin{sloppypar}
\samepage\noindent
A.~Heister,
S.~Schael
\nopagebreak
\begin{center}
\parbox{15.5cm}{\sl\samepage
Physikalisches Institut das RWTH-Aachen, D-52056 Aachen, Germany}
\end{center}\end{sloppypar}
\vspace{2mm}
\begin{sloppypar}
\noindent
R.~Barate,
I.~De~Bonis,
D.~Decamp,
C.~Goy,
\mbox{J.-P.~Lees},
E.~Merle,
\mbox{M.-N.~Minard},
B.~Pietrzyk
\nopagebreak
\begin{center}
\parbox{15.5cm}{\sl\samepage
Laboratoire de Physique des Particules (LAPP), IN$^{2}$P$^{3}$-CNRS,
F-74019 Annecy-le-Vieux Cedex, France}
\end{center}\end{sloppypar}
\vspace{2mm}
\begin{sloppypar}
\noindent
G.~Boix,
S.~Bravo,
M.P.~Casado,
M.~Chmeissani,
J.M.~Crespo,
E.~Fernandez,
\mbox{M.~Fernandez-Bosman},
Ll.~Garrido,$^{15}$
E.~Graug\'{e}s,
M.~Martinez,
G.~Merino,
R.~Miquel,$^{27}$
Ll.M.~Mir,$^{27}$
A.~Pacheco,
H.~Ruiz
\nopagebreak
\begin{center}
\parbox{15.5cm}{\sl\samepage
Institut de F\'{i}sica d'Altes Energies, Universitat Aut\`{o}noma
de Barcelona, E-08193 Bellaterra (Barcelona), Spain$^{7}$}
\end{center}\end{sloppypar}
\vspace{2mm}
\begin{sloppypar}
\noindent
A.~Colaleo,
D.~Creanza,
M.~de~Palma,
G.~Iaselli,
G.~Maggi,
M.~Maggi,
S.~Nuzzo,
A.~Ranieri,
G.~Raso,$^{23}$
F.~Ruggieri,
G.~Selvaggi,
L.~Silvestris,
P.~Tempesta,
A.~Tricomi,$^{3}$
G.~Zito
\nopagebreak
\begin{center}
\parbox{15.5cm}{\sl\samepage
Dipartimento di Fisica, INFN Sezione di Bari, I-70126
Bari, Italy}
\end{center}\end{sloppypar}
\vspace{2mm}
\begin{sloppypar}
\noindent
X.~Huang,
J.~Lin,
Q. Ouyang,
T.~Wang,
Y.~Xie,
R.~Xu,
S.~Xue,
J.~Zhang,
L.~Zhang,
W.~Zhao
\nopagebreak
\begin{center}
\parbox{15.5cm}{\sl\samepage
Institute of High Energy Physics, Academia Sinica, Beijing, The People's
Republic of China$^{8}$}
\end{center}\end{sloppypar}
\vspace{2mm}
\begin{sloppypar}
\noindent
D.~Abbaneo,
P.~Azzurri,
O.~Buchm\"uller,$^{25}$
M.~Cattaneo,
F.~Cerutti,
B.~Clerbaux,
H.~Drevermann,
R.W.~Forty,
M.~Frank,
F.~Gianotti,
T.C.~Greening,$^{29}$
J.B.~Hansen,
J.~Harvey,
D.E.~Hutchcroft,
P.~Janot,
B.~Jost,
M.~Kado,$^{27}$
P.~Mato,
A.~Moutoussi,
F.~Ranjard,
L.~Rolandi,
D.~Schlatter,
O.~Schneider,$^{2}$
G.~Sguazzoni,
W.~Tejessy,
F.~Teubert,
A.~Valassi,
I.~Videau,
J.~Ward
\nopagebreak
\begin{center}
\parbox{15.5cm}{\sl\samepage
European Laboratory for Particle Physics (CERN), CH-1211 Geneva 23,
Switzerland}
\end{center}\end{sloppypar}
\vspace{2mm}
\begin{sloppypar}
\noindent
F.~Badaud,
A.~Falvard,$^{22}$
P.~Gay,
P.~Henrard,
J.~Jousset,
B.~Michel,
S.~Monteil,
\mbox{J-C.~Montret},
D.~Pallin,
P.~Perret
\nopagebreak
\begin{center}
\parbox{15.5cm}{\sl\samepage
Laboratoire de Physique Corpusculaire, Universit\'e Blaise Pascal,
IN$^{2}$P$^{3}$-CNRS, Clermont-Ferrand, F-63177 Aubi\`{e}re, France}
\end{center}\end{sloppypar}
\vspace{2mm}
\begin{sloppypar}
\noindent
J.D.~Hansen,
J.R.~Hansen,
P.H.~Hansen,
B.S.~Nilsson,
A.~W\"a\"an\"anen
\begin{center}
\parbox{15.5cm}{\sl\samepage
Niels Bohr Institute, DK-2100 Copenhagen, Denmark$^{9}$}
\end{center}\end{sloppypar}
\vspace{2mm}
\begin{sloppypar}
\noindent
A.~Kyriakis,
C.~Markou,
E.~Simopoulou,
A.~Vayaki,
K.~Zachariadou
\nopagebreak
\begin{center}
\parbox{15.5cm}{\sl\samepage
Nuclear Research Center Demokritos (NRCD), GR-15310 Attiki, Greece}
\end{center}\end{sloppypar}
\vspace{2mm}
\begin{sloppypar}
\noindent
A.~Blondel,$^{12}$
G.~Bonneaud,
\mbox{J.-C.~Brient},
A.~Roug\'{e},
M.~Rumpf,
M.~Swynghedauw,
M.~Verderi,
\linebreak
H.~Videau
\nopagebreak
\begin{center}
\parbox{15.5cm}{\sl\samepage
Laboratoire de Physique Nucl\'eaire et des Hautes Energies, Ecole
Polytechnique, IN$^{2}$P$^{3}$-CNRS, \mbox{F-91128} Palaiseau Cedex, France}
\end{center}\end{sloppypar}
\vspace{2mm}
\begin{sloppypar}
\noindent
V.~Ciulli,
E.~Focardi,
G.~Parrini
\nopagebreak
\begin{center}
\parbox{15.5cm}{\sl\samepage
Dipartimento di Fisica, Universit\`a di Firenze, INFN Sezione di Firenze,
I-50125 Firenze, Italy}
\end{center}\end{sloppypar}
\vspace{2mm}
\begin{sloppypar}
\noindent
A.~Antonelli,
M.~Antonelli,
G.~Bencivenni,
G.~Bologna,$^{4}$
F.~Bossi,
P.~Campana,
G.~Capon,
V.~Chiarella,
P.~Laurelli,
G.~Mannocchi,$^{5}$
F.~Murtas,
G.P.~Murtas,
L.~Passalacqua,
\mbox{M.~Pepe-Altarelli},$^{24}$
P.~Spagnolo
\nopagebreak
\begin{center}
\parbox{15.5cm}{\sl\samepage
Laboratori Nazionali dell'INFN (LNF-INFN), I-00044 Frascati, Italy}
\end{center}\end{sloppypar}
\vspace{2mm}
\begin{sloppypar}
\noindent
A.~Halley,
J.G.~Lynch,
P.~Negus,
V.~O'Shea,
C.~Raine,$^{4}$
A.S.~Thompson
\nopagebreak
\begin{center}
\parbox{15.5cm}{\sl\samepage
Department of Physics and Astronomy, University of Glasgow, Glasgow G12
8QQ,United Kingdom$^{10}$}
\end{center}\end{sloppypar}
\vspace{2mm}
\begin{sloppypar}
\noindent
S.~Wasserbaech
\nopagebreak
\begin{center}
\parbox{15.5cm}{\sl\samepage
Department of Physics, Haverford College, Haverford, PA 19041-1392, U.S.A.}
\end{center}\end{sloppypar}
\vspace{2mm}
\begin{sloppypar}
\noindent
R.~Cavanaugh,
S.~Dhamotharan,
C.~Geweniger,
P.~Hanke,
G.~Hansper,
V.~Hepp,
E.E.~Kluge,
A.~Putzer,
J.~Sommer,
K.~Tittel,
S.~Werner,$^{19}$
M.~Wunsch$^{19}$
\nopagebreak
\begin{center}
\parbox{15.5cm}{\sl\samepage
Kirchhoff-Institut f\"ur Physik, Universit\"at Heidelberg, D-69120
Heidelberg, Germany$^{16}$}
\end{center}\end{sloppypar}
\vspace{2mm}
\pagebreak
\begin{sloppypar}
\noindent
R.~Beuselinck,
D.M.~Binnie,
W.~Cameron,
P.J.~Dornan,
M.~Girone,$^{1}$
N.~Marinelli,
J.K.~Sedgbeer,
J.C.~Thompson$^{14}$
\nopagebreak
\begin{center}
\parbox{15.5cm}{\sl\samepage
Department of Physics, Imperial College, London SW7 2BZ,
United Kingdom$^{10}$}
\end{center}\end{sloppypar}
\vspace{2mm}
\begin{sloppypar}
\noindent
V.M.~Ghete,
P.~Girtler,
E.~Kneringer,
D.~Kuhn,
G.~Rudolph
\nopagebreak
\begin{center}
\parbox{15.5cm}{\sl\samepage
Institut f\"ur Experimentalphysik, Universit\"at Innsbruck, A-6020
Innsbruck, Austria$^{18}$}
\end{center}\end{sloppypar}
\vspace{2mm}
\begin{sloppypar}
\noindent
E.~Bouhova-Thacker,
C.K.~Bowdery,
A.J.~Finch,
F.~Foster,
G.~Hughes,
R.W.L.~Jones,
M.R.~Pearson,
N.A.~Robertson
\nopagebreak
\begin{center}
\parbox{15.5cm}{\sl\samepage
Department of Physics, University of Lancaster, Lancaster LA1 4YB,
United Kingdom$^{10}$}
\end{center}\end{sloppypar}
\vspace{2mm}
\begin{sloppypar}
\noindent
K.~Jakobs,
K.~Kleinknecht,
G.~Quast,$^{6}$
B.~Renk,
\mbox{H.-G.~Sander},
H.~Wachsmuth,
C.~Zeitnitz
\nopagebreak
\begin{center}
\parbox{15.5cm}{\sl\samepage
Institut f\"ur Physik, Universit\"at Mainz, D-55099 Mainz, Germany$^{16}$}
\end{center}\end{sloppypar}
\vspace{2mm}
\begin{sloppypar}
\noindent
A.~Bonissent,
J.~Carr,
P.~Coyle,
O.~Leroy,
P.~Payre,
D.~Rousseau,
M.~Talby
\nopagebreak
\begin{center}
\parbox{15.5cm}{\sl\samepage
Centre de Physique des Particules, Universit\'e de la M\'editerran\'ee,
IN$^{2}$P$^{3}$-CNRS, F-13288 Marseille, France}
\end{center}\end{sloppypar}
\vspace{2mm}
\begin{sloppypar}
\noindent
F.~Ragusa
\nopagebreak
\begin{center}
\parbox{15.5cm}{\sl\samepage
Dipartimento di Fisica, Universit\`a di Milano e INFN Sezione di Milano,
I-20133 Milano, Italy}
\end{center}\end{sloppypar}
\vspace{2mm}
\begin{sloppypar}
\noindent
A.~David,
H.~Dietl,
G.~Ganis,$^{26}$
K.~H\"uttmann,
G.~L\"utjens,
C.~Mannert,
W.~M\"anner,
\mbox{H.-G.~Moser},
R.~Settles,
H.~Stenzel,
W.~Wiedenmann,
G.~Wolf
\nopagebreak
\begin{center}
\parbox{15.5cm}{\sl\samepage
Max-Planck-Institut f\"ur Physik, Werner-Heisenberg-Institut,
D-80805 M\"unchen, Germany\footnotemark[16]}
\end{center}\end{sloppypar}
\vspace{2mm}
\begin{sloppypar}
\noindent
J.~Boucrot,
O.~Callot,
M.~Davier,
L.~Duflot,
\mbox{J.-F.~Grivaz},
Ph.~Heusse,
A.~Jacholkowska,$^{22}$
J.~Lefran\c{c}ois,
\mbox{J.-J.~Veillet},
C.~Yuan
\nopagebreak
\begin{center}
\parbox{15.5cm}{\sl\samepage
Laboratoire de l'Acc\'el\'erateur Lin\'eaire, Universit\'e de Paris-Sud,
IN$^{2}$P$^{3}$-CNRS, F-91898 Orsay Cedex, France}
\end{center}\end{sloppypar}
\vspace{2mm}
\begin{sloppypar}
\noindent
G.~Bagliesi,
T.~Boccali,
L.~Fo\`{a},
A.~Giammanco,
A.~Giassi,
F.~Ligabue,
A.~Messineo,
F.~Palla,
G.~Sanguinetti,
A.~Sciab\`a,
R.~Tenchini,$^{1}$
A.~Venturi,$^{1}$
P.G.~Verdini
\samepage
\begin{center}
\parbox{15.5cm}{\sl\samepage
Dipartimento di Fisica dell'Universit\`a, INFN Sezione di Pisa,
e Scuola Normale Superiore, I-56010 Pisa, Italy}
\end{center}\end{sloppypar}
\vspace{2mm}
\begin{sloppypar}
\noindent
G.A.~Blair,
G.~Cowan,
M.G.~Green,
T.~Medcalf,
A.~Misiejuk,
J.A.~Strong,
\mbox{P.~Teixeira-Dias},
\mbox{J.H.~von~Wimmersperg-Toeller}
\nopagebreak
\begin{center}
\parbox{15.5cm}{\sl\samepage
Department of Physics, Royal Holloway \& Bedford New College,
University of London, Egham, Surrey TW20 OEX, United Kingdom$^{10}$}
\end{center}\end{sloppypar}
\vspace{2mm}
\begin{sloppypar}
\noindent
R.W.~Clifft,
T.R.~Edgecock,
P.R.~Norton,
I.R.~Tomalin
\nopagebreak
\begin{center}
\parbox{15.5cm}{\sl\samepage
Particle Physics Dept., Rutherford Appleton Laboratory,
Chilton, Didcot, Oxon OX11 OQX, United Kingdom$^{10}$}
\end{center}\end{sloppypar}
\vspace{2mm}
\begin{sloppypar}
\noindent
\mbox{B.~Bloch-Devaux},
P.~Colas,
S.~Emery,
W.~Kozanecki,
E.~Lan\c{c}on,
\mbox{M.-C.~Lemaire},
E.~Locci,
P.~Perez,
J.~Rander,
\mbox{J.-F.~Renardy},
A.~Roussarie,
\mbox{J.-P.~Schuller},
J.~Schwindling,
A.~Trabelsi,$^{21}$
B.~Vallage
\nopagebreak
\begin{center}
\parbox{15.5cm}{\sl\samepage
CEA, DAPNIA/Service de Physique des Particules,
CE-Saclay, F-91191 Gif-sur-Yvette Cedex, France$^{17}$}
\end{center}\end{sloppypar}
\vspace{2mm}
\begin{sloppypar}
\noindent
N.~Konstantinidis,
A.M.~Litke,
G.~Taylor
\nopagebreak
\begin{center}
\parbox{15.5cm}{\sl\samepage
Institute for Particle Physics, University of California at
Santa Cruz, Santa Cruz, CA 95064, USA$^{13}$}
\end{center}\end{sloppypar}
\vspace{2mm}
\begin{sloppypar}
\noindent
A.~Beddall,$^{30}$ 
C.N.~Booth,
S.~Cartwright,
F.~Combley,$^{4}$
M.~Lehto,
L.F.~Thompson
\nopagebreak
\begin{center}
\parbox{15.5cm}{\sl\samepage
Department of Physics, University of Sheffield, Sheffield S3 7RH,
United Kingdom$^{10}$}
\end{center}\end{sloppypar}
\vspace{2mm}
\begin{sloppypar}
\noindent
K.~Affholderbach,$^{28}$
A.~B\"ohrer,
S.~Brandt,
C.~Grupen,
A.~Ngac,
G.~Prange,
U.~Sieler
\nopagebreak
\begin{center}
\parbox{15.5cm}{\sl\samepage
Fachbereich Physik, Universit\"at Siegen, D-57068 Siegen,
 Germany$^{16}$}
\end{center}\end{sloppypar}
\vspace{2mm}
\begin{sloppypar}
\noindent
G.~Giannini
\nopagebreak
\begin{center}
\parbox{15.5cm}{\sl\samepage
Dipartimento di Fisica, Universit\`a di Trieste e INFN Sezione di Trieste,
I-34127 Trieste, Italy}
\end{center}\end{sloppypar}
\vspace{2mm}
\pagebreak
\begin{sloppypar}
\noindent
J.~Rothberg
\nopagebreak
\begin{center}
\parbox{15.5cm}{\sl\samepage
Experimental Elementary Particle Physics, University of Washington, Seattle, 
WA 98195 U.S.A.}
\end{center}\end{sloppypar}
\vspace{2mm}
\begin{sloppypar}
\noindent
S.R.~Armstrong,
K.~Berkelman,
K.~Cranmer,
D.P.S.~Ferguson,
Y.~Gao,$^{20}$
S.~Gonz\'{a}lez,
O.J.~Hayes,
H.~Hu,
S.~Jin,
J.~Kile,
P.A.~McNamara III,
J.~Nielsen,
Y.B.~Pan,
\mbox{J.H.~von~Wimmersperg-Toeller},
W.~Wiedenmann,
J.~Wu,
Sau~Lan~Wu,
X.~Wu,
G.~Zobernig
\nopagebreak
\begin{center}
\parbox{15.5cm}{\sl\samepage
Department of Physics, University of Wisconsin, Madison, WI 53706,
USA$^{11}$}
\end{center}\end{sloppypar}
\vspace{2mm}
\begin{sloppypar}
\noindent
G.~Dissertori
\nopagebreak
\begin{center}
\parbox{15.5cm}{\sl\samepage
Institute for Particle Physics, ETH H\"onggerberg, 8093 Z\"urich, Switzerland.}
\end{center}\end{sloppypar}
}
\footnotetext[1]{Also at CERN, 1211 Geneva 23, Switzerland.}
\footnotetext[2]{Now at Universit\'e de Lausanne, 1015 Lausanne, Switzerland.}
\footnotetext[3]{Also at Dipartimento di Fisica di Catania and INFN Sezione di
 Catania, 95129 Catania, Italy.}
\footnotetext[4]{Deceased.}
\footnotetext[5]{Also Istituto di Cosmo-Geofisica del C.N.R., Torino,
Italy.}
\footnotetext[6]{Now at Institut f\"ur Experimentelle Kernphysik, Universit\"at Karlsruhe, 76128 Karlsruhe, Germany.}
\footnotetext[7]{Supported by CICYT, Spain.}
\footnotetext[8]{Supported by the National Science Foundation of China.}
\footnotetext[9]{Supported by the Danish Natural Science Research Council.}
\footnotetext[10]{Supported by the UK Particle Physics and Astronomy Research
Council.}
\footnotetext[11]{Supported by the US Department of Energy, grant
DE-FG0295-ER40896.}
\footnotetext[12]{Now at Departement de Physique Corpusculaire, Universit\'e de
Gen\`eve, 1211 Gen\`eve 4, Switzerland.}
\footnotetext[13]{Supported by the US Department of Energy,
grant DE-FG03-92ER40689.}
\footnotetext[14]{Also at Rutherford Appleton Laboratory, Chilton, Didcot, UK.}
\footnotetext[15]{Permanent address: Universitat de Barcelona, 08208 Barcelona,
Spain.}
\footnotetext[16]{Supported by the Bundesministerium f\"ur Bildung,
Wissenschaft, Forschung und Technologie, Germany.}
\footnotetext[17]{Supported by the Direction des Sciences de la
Mati\`ere, C.E.A.}
\footnotetext[18]{Supported by the Austrian Ministry for Science and Transport.}
\footnotetext[19]{Now at SAP AG, 69185 Walldorf, Germany.}
\footnotetext[20]{Also at Department of Physics, Tsinghua University, Beijing, The People's Republic of China.}
\footnotetext[21]{Now at D\'epartement de Physique, Facult\'e des Sciences de Tunis, 1060 Le Belv\'ed\`ere, Tunisia.}
\footnotetext[22]{Now at Groupe d' Astroparticules de Montpellier, Universit\'e de Montpellier II, 34095 Montpellier, France.}
\footnotetext[23]{Also at Dipartimento di Fisica e Tecnologie Relative, Universit\`a di Palermo, Palermo, Italy.}
\footnotetext[24]{Now at CERN, 1211 Geneva 23, Switzerland.}
\footnotetext[25]{Now at SLAC, Stanford, CA 94309, U.S.A.}
\footnotetext[26]{Now at INFN Sezione di Roma II, Dipartimento di Fisica, Universit\'a di Roma Tor Vergata, 00133 Roma, Italy.} 
\footnotetext[27]{Now at LBNL, Berkeley, CA 94720, U.S.A.}
\footnotetext[28]{Now at Skyguide, Swissair Navigation Services, Geneva, Switzerland.}
\footnotetext[29]{Now at Honeywell, Phoenix AZ, U.S.A.}
\footnotetext[30]{Now at Department of Engineering Physics,
University of Gaziantep, Turkey.}
\setlength{\parskip}{\saveparskip}
\setlength{\textheight}{\savetextheight}
\setlength{\topmargin}{\savetopmargin}
\setlength{\textwidth}{\savetextwidth}
\setlength{\oddsidemargin}{\saveoddsidemargin}
\setlength{\topsep}{\savetopsep}
\normalsize
\newpage
\pagestyle{plain}
\setcounter{page}{1}

\pagenumbering{arabic}


\section{Introduction}

The description of the hadronization process in QCD is deeply connected 
with the confinement property and requires non-perturbative methods. 
The precise measurement of momentum spectra of 
identified particles in the clean 
environment of $\epem$ annihilation into hadrons may improve the understanding 
of hadronization. Meanwhile these measurements are necessary to test and 
tune the phenomenological models used to describe the hadronization: each 
of these models has free parameters which must be determined from comparison 
with data \cite{reviews}.


Resonant states and their
dynamics are more closely related to the original partons than
pions. The $\eta$ and $\omega(782)$ mesons are copiously produced
in hadronic events and thus well suited to a study of hadronization. 

In this investigation a measurement of the inclusive momentum
distributions of the $\eta$ and $\omega$ mesons obtained through 
their $\pip \pim \piz$ decays is presented,
which improves on the previous ALEPH results \cite{NVMP} with higher 
statistics and reduced systematic effects. It also complements the recent 
publication concerning the $\gamma\gamma$ decay of the 
$\eta$ meson \cite{NEWETA}. 
Studies by the OPAL and L3 Collaborations can be found in 
\cite{OPAL,L3neweta,L3}.


The measurement of the partial width of leptonic decay rates of the
light vector mesons is a good test both for the quark assignments 
($\sum a_q \qqbar$ with $q= {\mathrm{u}},{\mathrm{d}},{\mathrm{s}}$) 
in the vector meson and the quark charges. 
With $\rho^0 = (\uubar - \ddbar)/\sqrt{2}$,
$\omega = (\uubar + \ddbar)/\sqrt{2}$ and $\phi = \ssbar$ the
partial decay width of a vector meson V to lepton pairs
($\ell$=e$^{\pm}$, $\mu^{\pm}$)
can be calculated using the
Van Royen-Weisskopf formula \cite{royen}:
\begin{displaymath}
\Gamma_{\ell} (\mathrm{V \rightarrow \ell^+\ell^-}) =
\frac{16\pi \alpha^2 Q^2}
     { m_{\mathrm{V}}^2 } | \psi(0) | ^2 \ ,
\end{displaymath}
where $Q^2 = ( \sum a_q Q_q ) ^2$ is the square of the flavour-weighted
sum of the charges $Q_q$ of the quarks 
($q= {\mathrm{u}},{\mathrm{d}},{\mathrm{s}}$), 
$\psi(0)$ is the amplitude of the $\qqbar$ wavefunction
at the origin, and $m_{\mathrm{V}}$ is the meson mass.
The expected ratio of the leptonic widths is
$\Gamma_{\ell} (\rho^0) : \Gamma_{\ell} (\omega) : \Gamma_{\ell} (\phi) 
= 9:1:2$ in
agreement with the widths and branching ratios measured as summarized in
\cite{PDG98} for the decay to $\epem$. While
$\Gamma_{\ell} (\rho^0)$ and $\Gamma_{\ell} (\phi)$
are in agreement with the prediction based on the Van Royen-Weisskopf 
formula and lepton
universality also for the muonic decay, only an upper limit \cite{PDG98} 
so far has been established 
for the decay $\omega \rightarrow \mu^+\mu^-$ 
(Table~\ref{tab-BR}). The muonic branching
ratio of the $\omega $ is measured here for the first time.

\begin{table}[hptb]
\begin{center}
\caption{\small Widths and leptonic branching ratios of the light vector
mesons V \cite{PDG98}\label{tab-BR}}
\vspace{3mm}
\begin{tabular}{|l|r|r|r|}
\hline
      & $\Gamma$ [MeV]
                 & BR$(\mathrm{V} \rightarrow \epem)$
                 & BR$(\mathrm{V} \rightarrow \mu^+ \mu^-)$  \\
\hline
\hline
$\rho^0$   &    150.2$\pm$0.8
  &  $(4.49 \pm 0.22)\times 10^{-5}$  &    $(4.60 \pm 0.28)\times 10^{-5}$\\
\hline
$\omega$   &    8.44$\pm$0.09
  &  $(7.07 \pm 0.19)\times 10^{-5}$  &    (CL=90\%) $< 1.8 \times10^{-4}$\\
\hline
$\phi$     &    4.46$\pm$0.03
  &  $(2.91 \pm 0.07)\times 10^{-4}$  &    $(3.7 \pm 0.5) \times 10^{-4}$\\
\hline
\end{tabular}
\end{center}
\end{table}

After a brief description of the ALEPH detector, the selection of hadronic 
events is detailed in Section 3 and the particle identification in 
Section 4. 
The improved measurement of inclusive $\eta$ and $\omega$ production
in the $\pi^{+} \pi^{-} \pi^{0}$ channel with a large data sample 
of about 4 million hadronic Z decays is presented in Section \ref{3PI}.
The measurement of the branching ratio $\omega \rightarrow \mu^+ \mu^-$ 
is presented in Section \ref{2MU}, followed by a summary and conclusions.

\section{The ALEPH Detector}
The ALEPH detector is described in detail elsewhere 
\cite{DETECTOR,PERFORMANCE}. Charged particles are measured over
the polar angle range $|\mathrm{cos} \, \theta| < 0.85$ and 
$|\mathrm{cos} \, \theta| < 0.69$ by the two layers of the silicon 
vertex detector (VDET). This is surrounded by a 
cylindrical inner drift chamber, and a large cylindrical time
projection chamber (TPC) which measures up to 21 three-dimensional
space points per track. 
A particle's energy loss is sampled in the TPC by up to $338$ 
wires and 21 pads. The tracking detectors 
are immersed in a magnetic field of $1.5$ $\mathrm{T}$ 
and provide a momentum resolution of 
$\delta p_t/p_t = 0.0006 p_t
\oplus 0.005$ ($p_t$ in GeV/$c$).
The TPC is surrounded by an electromagnetic calorimeter (ECAL) 
of lead-proportional tube construction, 
which covers the angular range $|\mathrm{cos} \, \theta| < 0.98$ and 
has a thickness of $22$ radiation lengths. It is finely segmented 
in projective towers of approximately $0.9^{\circ}$ by $0.9^{\circ}$ 
providing an angular resolution of 
$\sigma_{\theta,\phi} = 2.5/\sqrt{E} + 0.25$ ($E$ in $\mathrm{GeV}$; 
$\sigma_{\theta,\phi}$ in $\mathrm{mrad}$). 
The energy resolution is 
$\sigma_{E}/E = 0.18/\sqrt{E} + 0.009$ ($E$ in $\mathrm{GeV}$) 
for isolated showers.
The hadron calorimeter (HCAL) uses the iron return yoke as an absorber, 
for a total of 7.5 hadronic interaction lengths. The iron is interleaved 
by 23 layers of streamer tubes, which provide a two-dimensional
measurement of muon tracks and a
view of the hadronic shower development.
The HCAL is used in conjunction with the muon chambers,
which are two double-layers of streamer tubes with three-dimensional readout,
and the tracking detectors to identify muons. The calorimeters
and the muon chambers cover nearly the
entire $4\pi$ solid angle.

\section{\label{EvSe}Event Selection}
For the event selection, good tracks are defined as originating close to 
the interaction point (with transverse impact parameter 
$|d_0| < 2$ $\mathrm{cm}$ and longitudinal impact parameter 
$|z_0| < 5$ $\mathrm{cm}$), having at least 
4 TPC hits, a polar angle in the range $20^{\circ} < \theta < 160^{\circ}$, 
and a transverse momentum $p_t > 200 \unit{MeV}/c$.
Four million hadronic $\zed$ decays are selected by requiring at least five 
good tracks. The total energy carried by all good tracks
is required to exceed $15$ $\mathrm{GeV}$ and the 
sphericity axis must be in the range $35^{\circ} < \theta < 145^{\circ}$.
With these cuts, a sample of 3.0 million events is selected.
The background to these events arises from tau pairs 
and two-photon events and is estimated to be less than $0.4 \%$ \cite{HADR}. 

For the measurement of the branching ratio BR($\omega \rightarrow \mm$)
additional event cuts are applied to reject heavy flavour events and events
with missing energy: events from
${\mathrm Z} \rightarrow {\mathrm{b} \mathrm{\bar{b}}}$ and
${\mathrm Z} \rightarrow {\mathrm{c} \mathrm{\bar{c}}}$
are rejected by a lifetime tag \cite{b-tag} keeping about 95\% 
of light flavours. A further reduction is obtained
by a cut on the missing energy corrected with hemisphere masses \cite{emiss}, 
$E_{\mathrm {miss}} < 6\unit{GeV}$, in the $\omega$ hemisphere.

For the purpose of comparing with models and as a means 
of measuring the detector acceptance,
samples of events generated with the {\sc Jetset 7.4} Monte Carlo
\cite{JETSET},
modified with {\sc Dymu3} {\cite{DYMU3}} for electromagnetic radiative 
effects and improved bottom and charm decay tables, were passed through 
a full detector simulation and reconstruction program. 
The generator was tuned to describe the ALEPH data using the inclusive charged 
particle and event shape distributions
{\cite{TUNE}}.
For model comparison and as a systematic check for the
extrapolation into the unmeasured region the measured spectra
were compared to those of {\sc Herwig 5.9} {\cite{HERWIG}}. 
The model parameters were tuned using ALEPH data in the
same manner as mentioned above.
For the measurement of the branching ratio BR($\omega \rightarrow \mm$)
a sample of events was generated with the {\sc Jetset 7.4} Monte Carlo, 
but with the muonic branching ratio of the
$\omega$ meson BR$(\omega \rightarrow \mu^+ \mu^-)$ set to 100\%. After 
a full detector simulation and reconstruction, the 
generated events were subjected to the same selection and analysis
chain as the data. 

\section{\label{PID}Particle Identification}
The $\eta$ and $\omega$ mesons are produced and decay at or near the 
primary interaction point. To reduce background from charged particles
produced far from the interaction point, e.g., from  K$^0_{\mathrm S}$ 
decays or photon conversions, 
track candidates are required to be reconstructed as good tracks
(as discussed above) with tighter cuts on the impact parameters 
($|d_0| < 0.5\unit{cm}$ and $|z_0| < 3.0\unit{cm}$) and 
the transverse momentum ($p_t > 250\unit{MeV}/c$). 

For the measurement of the branching ratio BR($\omega \rightarrow \mm$)
additional cuts are applied to identify and select muons above
the background of mainly charged pions. One VDET hit is required. 
For the identification of muons the measured energy loss 
in the TPC, ${\mathrm d}E/{\mathrm d}x_{\mathrm{meas}}$ 
should be consistent with the expectation 
for a muon, $({\mathrm d}E/{\mathrm d}x_{\mathrm{meas}}-
{\mathrm d}E/{\mathrm d}x_{\mathrm{exp}}) / \sigma_{\mathrm{exp}} 
> -2$, where ${\mathrm d}E/{\mathrm d}x_{\mathrm{exp}}$ and 
$\sigma_{\mathrm{exp}}$ are the expected energy loss and its resolution. 
The pattern in HCAL and the muon chambers should match with the 
pattern expected for penetrating muons, if the momentum is larger than 
$2.4\unit{GeV}/c$. For lower momenta, $1.5\unit{GeV}/c < p < 2.4\unit{GeV}/c$, 
the number of fired planes within the last ten HCAL planes divided 
by the number of all fired planes should be between 0.35 and 
0.75 \cite{muonHFid,muonid}. 

For the selection of neutral pions two-photon
invariant mass spectra are formed.
The photon energy is estimated from the energy collected in the 
four central ECAL towers of a cluster, correcting to the full energy from a 
parametrization of the shower shape for a single photon in the calorimeter. 
While the energy resolution is degraded to
$\sigma_{E}/E \approx 0.25/\sqrt{E}$ 
($E$ in $\mathrm{GeV}$) with this technique, hadronic background and 
clustering effects are reduced.
Photon candidates are only accepted if the estimated energy is
greater than $0.8$ $\mathrm{GeV}$.
Neutral pion candidates are accepted if the energy of the photon pair is 
less than $16$ $\mathrm{GeV}$ (to remove large uncertainties in the
acceptance correction at high energy).
The invariant mass of the photon pair must also be within $\pm 3 \sigma$ of
the expected mass. The $\pi^0$ energy resolution is improved
by constraining the mass of the $\pi^0$ candidates to $135$ $\mathrm{MeV}/c^2$.
The poor purity at low momentum, due to the large multiplicity
of low energy photons giving rise to a large combinatorial background,
is improved by a ``ranking'' method: 
candidates that share photons with other candidates are ranked
in an order determined by a $\piz$ estimator. 
The estimator $R$ is calculated for each candidate from
the photon pair opening angle $\theta_{12}$ and the $\chi^2$ from
the mass constraint:
\begin{equation}
      R = \theta_{12} ( 1.0 + 0.1 \chi^2 ). 
\end{equation}
    All $\piz$ candidates sharing photons with other candidates are 
    removed except the one with the smallest value of $R$.


\section{\label{3PI}
        Inclusive Production of \boldmath $\eta$ and \boldmath $\omega$}

\subsection{\label{CH_OMEGA_EXTRACT}
            Extraction of the \boldmath $\eta$ and \boldmath $\omega$ Signals}

The $\eta$ and $\omega$ cross sections are extracted from invariant 
mass distributions in the $\pip \pim \piz$ decay mode.
All charged tracks are assigned the charged pion mass; for neutral 
pion candidates the kinematic
refit constrains the mass to the $\piz$ mass.
Data are analysed in six intervals of scaled momentum
$x_p = p_{\mathrm {meson}}/p_{\mathrm {beam}}$, where $p_{\mathrm {beam}}$ is 
the LEP beam momentum. 
Only candidates satisfying $x_p > 0.10$ for the $\eta$ and 
$x_p > 0.05$ for the $\omega$ are considered, 
as the signal to background ratio is too small
for lower momenta.

\begin{figure}[t]
\begin{center}
\epsfig{file=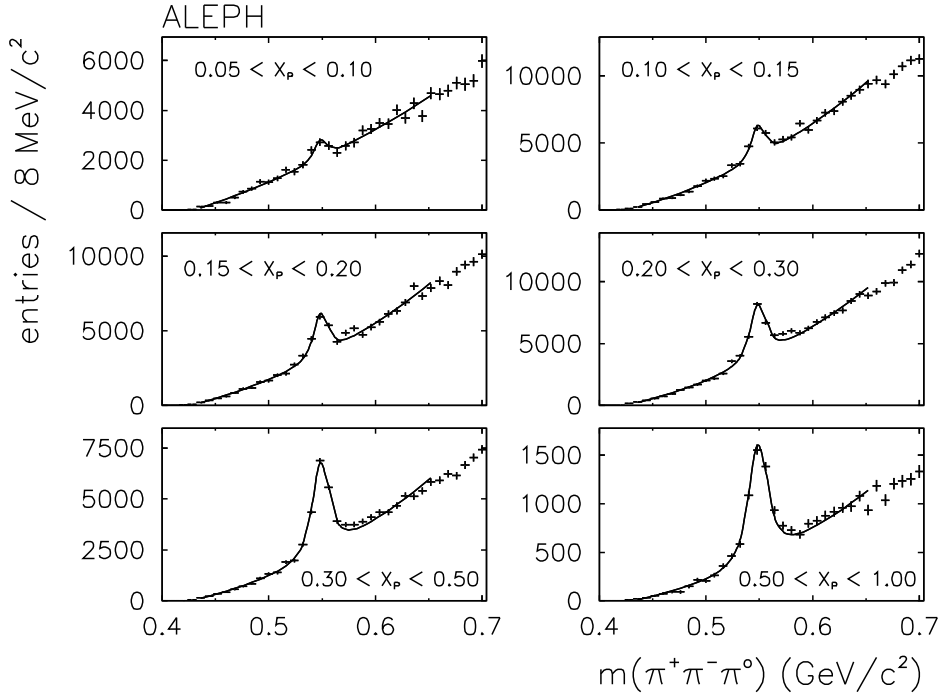,height=15cm}
\end{center}
\vspace{-56mm}
\caption[\small Fits made to the invariant mass spectra of 
        data.]{\label{ETA_DAFITS}
        {\small Fits in the $\eta$ region made to the invariant mass 
         spectra of data (like-sign subtracted).}}
\end{figure}

\begin{figure}[t]
\begin{center}
\epsfig{file=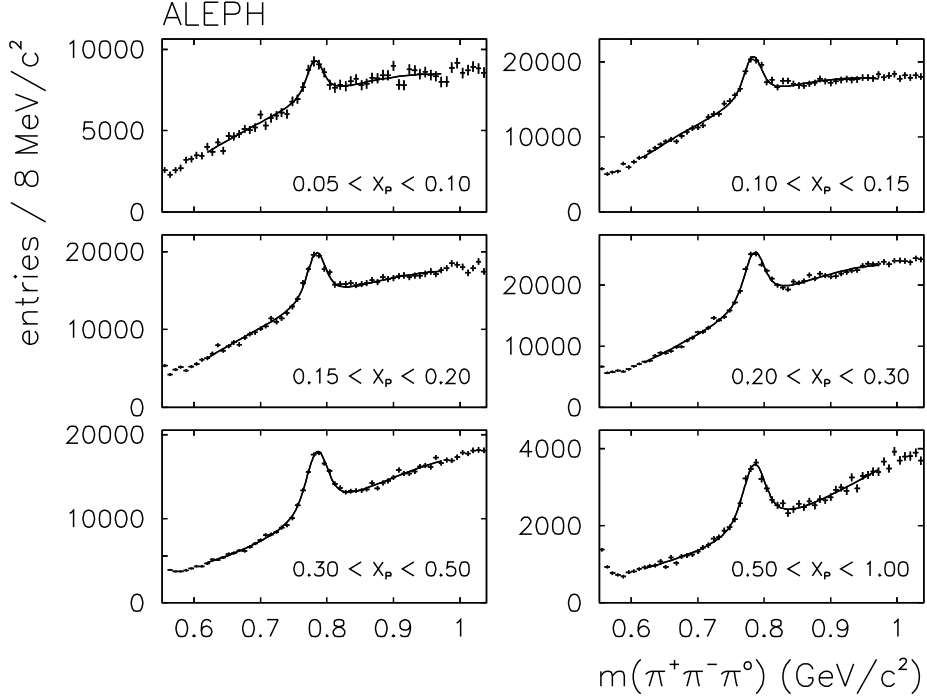,height=15cm}
\end{center}
\vspace{-56mm}
\caption[\small Fits made to the invariant mass spectra of 
        data.]{\label{OME_DAFITS}
        {\small Fits in the $\omega$ region made to the invariant mass 
         spectra of data (like-sign subtracted).}}
\end{figure}

The invariant mass spectra are fitted
as the sum of a background and a signal function.
Subtracting the like-sign spectra from the unlike-sign spectra
allows the backgrounds to be represented simply by a third order
polynomial. The signal function is empirically parametrized as
the sum of three Gaussian functions; 
the relative widths and normalizations of the three Gaussians
are determined from the Monte Carlo and are fixed with the overall 
normalization left free.
The results of fits to the data for the $\eta$ and $\omega$
are shown in Figs.~\ref{ETA_DAFITS} and \ref{OME_DAFITS},  
respectively.
The fit to the first momentum interval for the $\eta$ 
yields a large ($\approx$50\%) uncertainty, therefore this
momentum interval is not considered further.

%
%

\subsection{\label{OMEGA_RATES}
               Measured Rates and Differential Cross-Section}

The production rate per event, $R$, is calculated for each
momentum interval by correcting the fitted signal $S$ for the
reconstruction efficiency $\varepsilon$ and normalizing to one event.
The rate is corrected for the branching ratio
BR$(\eta \rightarrow \pip \pim \piz) = 0.230 \pm 0.004$, 
BR$(\omega \rightarrow \pip \pim \piz) = 0.888 \pm 0.007$, 
and BR$(\piz \rightarrow \gamma \gamma) = 0.9880 \pm 0.0003$ \cite{PDG98} 
to give the rate for all decay modes.
The calculation is as follows:
\begin{equation}
  R = \frac{S}{N} \frac{1}{\varepsilon} \frac{1}{\mathrm{BR}}
\end{equation}
where $N$ is the total number of hadronic events in data and $\varepsilon$ is 
the efficiency. 
$N$ is obtained as $N=N_{\mathrm {gen}} \cdot N_{\mathrm {acc}}^{\mathrm d}/ 
N_{\mathrm {acc}}^{\mathrm {MC}}$, 
where $N_{\mathrm {gen}}$ is the number of generated events in the 
Monte Carlo, 
while $N_{\mathrm {acc}}^{\mathrm d}$ and $N_{\mathrm {acc}}^{\mathrm {MC}}$ 
are the numbers of accepted events after 
preselection in data and Monte Carlo, respectively. 
The efficiency $\varepsilon$ is defined as 
$\varepsilon =  n_{\mathrm {rec}}/n_{\mathrm {gen}}$, 
where $n_{\mathrm {rec}}$ is the number of reconstructed 
$\eta$ ($\omega$) mesons in the
Monte Carlo, and 
$n_{\mathrm {gen}}$
is the number of generated $\eta$ ($\omega$) mesons in the
\MC (before event selection cuts).

The implementation of the $\eta \rightarrow \pip \pim \piz$ decay 
in {\sc Jetset} is according to phase space, 
while for the $\omega \rightarrow \pip \pim \piz$ decay it is according 
to the correct matrix element. For this analysis 
the Monte Carlo was reweighted for the $\eta$ decay by parametrizing the 
decay transition probability $\lambda_{\eta}$ as 
$\lambda_{\eta} \propto 1-T^*_0/T^*_{0,{\mathrm{max}}}$ as suggested in 
\cite{OPAL}, where $T^*_0$ is the kinetic energy of the neutral pion 
in the $\eta$ rest frame and $T^*_{0,{\mathrm{max}}}$ is its maximum possible 
value. This form is consistent with $\lambda_{\eta} \propto 1 + 2 \alpha y$ 
proposed in \cite{carpenterME} and the measurements in 
\cite{carpenterME,othersME}, where $y=(3T^*_0/Q)-1$, $Q$ is the mass 
difference between the $\eta$ and its daughters, and 
$\alpha =-0.47 \pm 0.04$. 
The $\eta$ efficiency is taken as the average of the results obtained 
with the two parametrizations.

The overall efficiency in the measured momentum region is 15.3\% for the
$\eta$ and 11.2\% for the $\omega$.
The measurements cover only the momentum region 
$x_p > 0.10$ for the $\eta$ and $x_p > 0.05$ for the $\omega$.
To estimate the total production rate the 
measured rates are extrapolated to $x_p = 0$ using the fragmentation
function in the Monte Carlo.
The fraction of $\eta$ mesons for $x_p > 0.10$ is 29.56\%.
For the $\omega$ meson the fraction for $x_p > 0.05$ is 58.8\%.

\subsection{\label{CH_OMEGA_SYS}Systematic Uncertainties}

The largest systematic uncertainties are determined to stem from
the photon energy cut, the impact parameter cut, fit range
and signal width, description of the neutral pion spectrum, 
and the uncertainty in the $\eta$ and $\omega$ branching ratios.
Systematic uncertainties are determined by varying cuts and taking
the largest change in the calculated rate as the systematic error.
Additionally, a systematic error is assigned representing the
uncertainty in the extrapolation of the measured rate to $x_p$ = 0.
Details of the systematic checks are given below.

A cut of $0.8$ $\mathrm{GeV}$ is applied on the energy of photons.
This cut is varied by $\pm 0.1$ $\mathrm{GeV}$.
A $d_0$ cut of $0.5$ $\mathrm{cm}$ is applied to remove tracks
originating from
decays of particles at some distance away from the primary interaction point.
As this cut is tight 
a check is performed to determine any associated systematic uncertainty.
This cut is decreased to $0.3$ $\mathrm{cm}$ and increased to
$1.0$ $\mathrm{cm}$.
Other cuts on charged track parameters do not
have significant systematic uncertainties associated with them.

In the fitting procedure mass spectra are fitted six times
with a different fit range.
The value for the extracted signal is seen to vary;  
a systematic error is assigned to this, calculated as the
standard deviation of the six results.
For the $\eta$, significant variations in extracted signals
occur for different choices of the background function, and 
a systematic error is assigned for this fitting uncertainty.
The fitting procedure is repeated with the width of the 
signal function as a free parameter, and the variation in the number
of extracted mesons is taken as a systematic error.
To check the effect of fixing the mass in the fitting procedure, the
procedure is repeated with the mass as a free parameter in the fit.
The fitted masses are found to be stable and no significant
variation in the extracted number of mesons was found,  
therefore no systematic error is assigned.
Uncertainties in the meson reconstruction efficiency arise from the 
modelling of the $\pi^0$ spectrum. 
It has been checked that bin-to-bin migration effects
are below 1\% for all but the highest $x_p$ bin. Even there the
estimated effect is smaller than other relevant systematic uncertainties.

For the $\eta$, the effect of the matrix element
proposed by \cite{OPAL} and \cite{carpenterME} which is not included in
{\sc Jetset} is corrected for by reweighting {\sc Jetset}
to agree with proposed parametrizations. Although little effect on the
momentum spectra is seen for high $\eta$ momenta, the effect at low
momenta is large (23\% in the lowest measured momentum interval).
As a correction, the mean of the two weighting schemes is applied to
the measured rates, and the difference in the two schemes in each momentum
interval is taken as the systematic error.

Statistical and systematic errors are summarized in 
Tables~\ref{TAB_ETA_ERRORS} and \ref{TAB_OME_ERRORS} for the
$\eta$ and $\omega$ measurements, respectively. 
For the systematic errors the individual errors
from each source are shown for each measured momentum interval (1 to 6).
The error for the total measured momentum range,
calculated for each error source, is taken as
the sum of the errors in each interval 
weighted by the rate in each interval.
The calculations take into consideration whether the errors
are correlated or not.
The statistical errors are taken from the fitting procedure.
The last line shows the final error for each momentum interval.

The extrapolation of the measured rates to $x_p = 0$ relies on
the \MC to give the correct scale factor for the extrapolation.
For the extrapolation the shape of the fragmentation function
in {\sc Jetset 7.4} is used.
To estimate the uncertainty in the scale factor the calculation is
repeated for the {\sc Herwig} Monte Carlo. 
For the $\eta$, extrapolating from $x_p$ = 0.10,
the difference, 5.8\%, between {\sc Herwig 5.9} and {\sc Jetset 7.4}
is taken as the systematic uncertainty in the extrapolation.
For the $\omega$, extrapolating from $x_p$ = 0.05,
the difference, 2.4\%, between {\sc Herwig 5.9} and {\sc Jetset 7.4}
is taken as the systematic uncertainty in the extrapolation.

\begin{table}[phtb]
\begin{center}
\caption[\small Systematic and statistical errors for the $\omega$ rate.]
        {\label{TAB_ETA_ERRORS}
        {\small Systematic and statistical errors on the $\eta$ 
        production rate in 
          each measured momentum interval. All values are expressed in 
          percent.}}
\vspace{3mm}
\begin{tabular}{|c||r||r|r|r|r|r|r|}
\hline
               & \multicolumn{7}{c|}{measured $x_p$ interval}              \\ \cline{2-8}
 Source of error ($\eta$)    &   all   & $1$ & $2$ & $3$ & $4$ & $5$ & $6$ \\ \hline
\hline 
 Photon energy cut           & 4.5   & - &12.1 & 6.7 & 3.4 & 1.2 & 1.4   \\ \hline
 Track selection             & 1.1   & - & 2.5 & 3.2 & 0.4 & 1.7 & 1.7   \\ \hline
 Fit range and background    & 2.0   & - & 5.0 & 3.3 & 3.0 & 1.4 & 2.0   \\ \hline
 Fit width                   & 1.5   & - & 3.2 & 4.5 & 0.4 & 2.6 & 1.4   \\ \hline
 $\piz$ spectrum             & 2.8   & - & 2.8 & 2.8 & 2.5 & 2.9 & 4.8   \\ \hline
 $\eta$ branching ratio      & 2.2   & - & 2.2 & 2.2 & 2.2 & 2.2 & 2.2   \\ \hline
 $\eta$ matrix element       & 2.3   & - & 3.9 & 2.2 & 1.4 & 1.1 & 0.3   \\ \hline
\hline 
 Total systematic error      & 6.8   & - &14.7 &10.2 & 5.8 & 5.2 & 6.2   \\ \hline
 Statistical error           & 3.1   & - & 7.9 & 5.6 & 3.5 & 2.3 & 3.0   \\ \hline
\hline 
 Total error                 & 7.5   & - &16.7 &11.6 & 6.8 & 5.7 & 6.9   \\ 
\hline 
\end{tabular}
\end{center}
\end{table}

\begin{table}[phtb]
\begin{center}
\caption[\small Systematic and statistical errors for the $\omega$ rate.]
        {\label{TAB_OME_ERRORS}
        {\small Systematic and statistical errors on the $\omega$ 
        production rate in 
          each measured momentum interval. All values are expressed in 
          percent.}}
\vspace{3mm}
\begin{tabular}{|c||r||r|r|r|r|r|r|}
\hline
               & \multicolumn{7}{c|}{measured $x_p$ interval}               \\ \cline{2-8}
 Source of error ($\omega$) &   all    & $1$ & $2$ & $3$ & $4$ & $5$ & $6$  \\ \hline
\hline 
 Photon energy cut         & 4.1  &10.6 & 2.2 & 2.8 & 1.7 & 0.1 & 1.8  \\ \hline
 Track selection           & 2.6  & 6.7 & 1.9 & 1.2 & 1.1 & 0.5 & 1.1  \\ \hline
 Fit range                 & 1.0  & 1.7 & 3.2 & 2.2 & 2.0 & 0.3 & 1.1  \\ \hline
 Fit width                 & 0.8  & 1.0 & 3.3 & 1.8 & 0.4 & 0.2 & 0.5  \\ \hline
 $\piz$ spectrum           & 2.7  & 2.5 & 2.8 & 2.8 & 2.5 & 2.9 & 4.8  \\ \hline
 $\omega$ branching ratio  & 0.7  & 0.7 & 0.7 & 0.7 & 0.7 & 0.7 & 0.7  \\ \hline
\hline 
 Total systematic error    & 5.7  &13.0 & 6.2 & 5.1 & 3.9 & 3.0 & 5.4  \\ \hline
 Statistical error         & 3.2  & 7.8 & 4.2 & 3.3 & 2.8 & 2.0 & 2.9  \\ \hline
\hline 
 Total error               & 6.5  &15.2 & 7.5 & 6.1 & 4.8 & 3.6 & 6.1  \\ 
\hline 
\end{tabular}
\end{center}
\end{table}

\subsection{\label{CH_RESULTS}Results}
              
\begin{figure}[p]
\vspace{-10mm}
\begin{center}
\epsfig{file=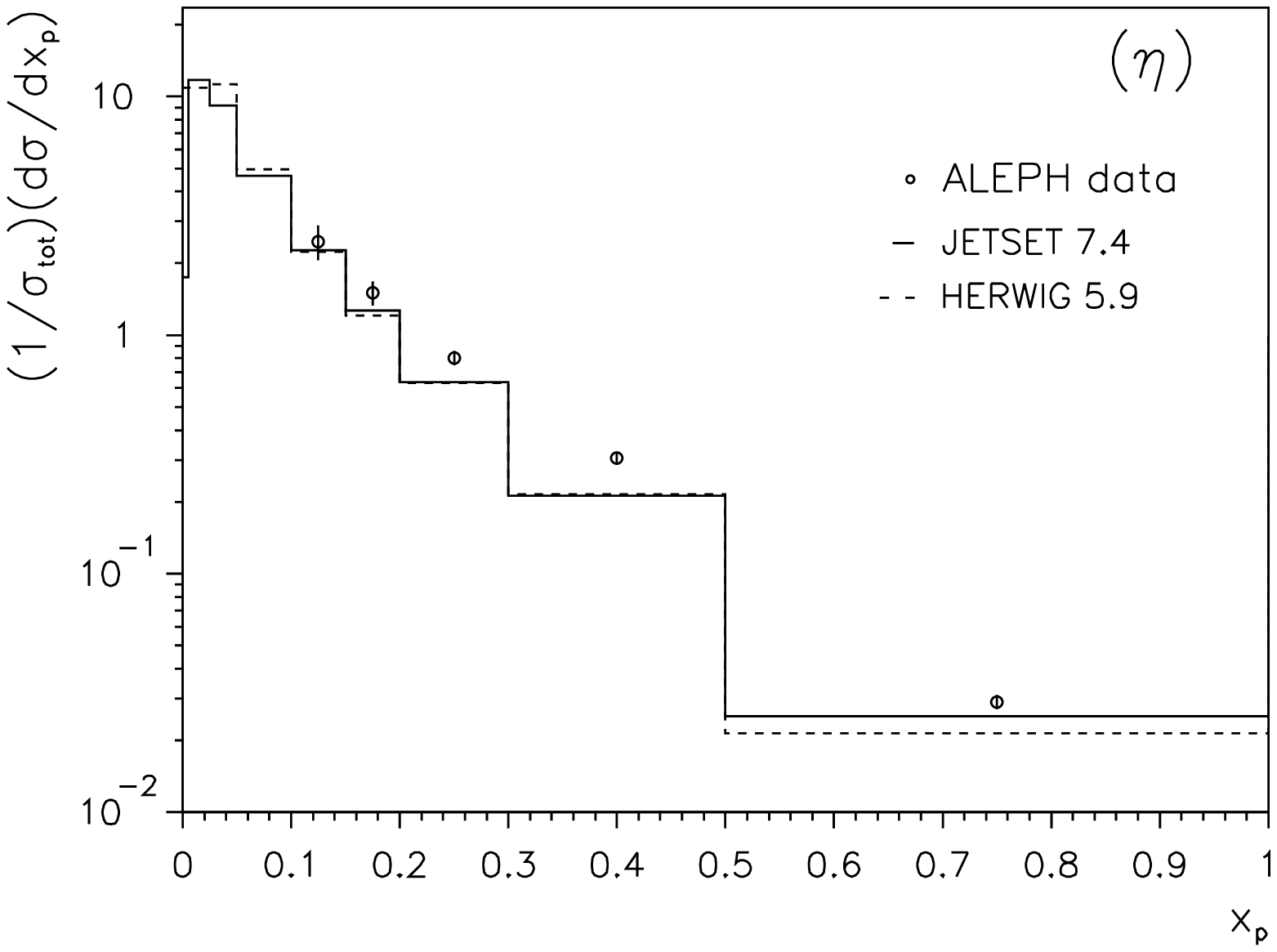,height=12cm}
\end{center}
\vspace{-50mm}
\caption[\small Measured $\eta$ cross section in comparison with 
        MC predictions.]
        {\label{ETA_RESULTS}
        {\small Measured differential cross sections for the $\eta$
          in comparison with \MC predictions. The errors shown are the 
          quadratic sum of statistical and systematic contributions. }}
\end{figure}

\begin{figure}[p]
\vspace{-10mm}
\begin{center}
\epsfig{file=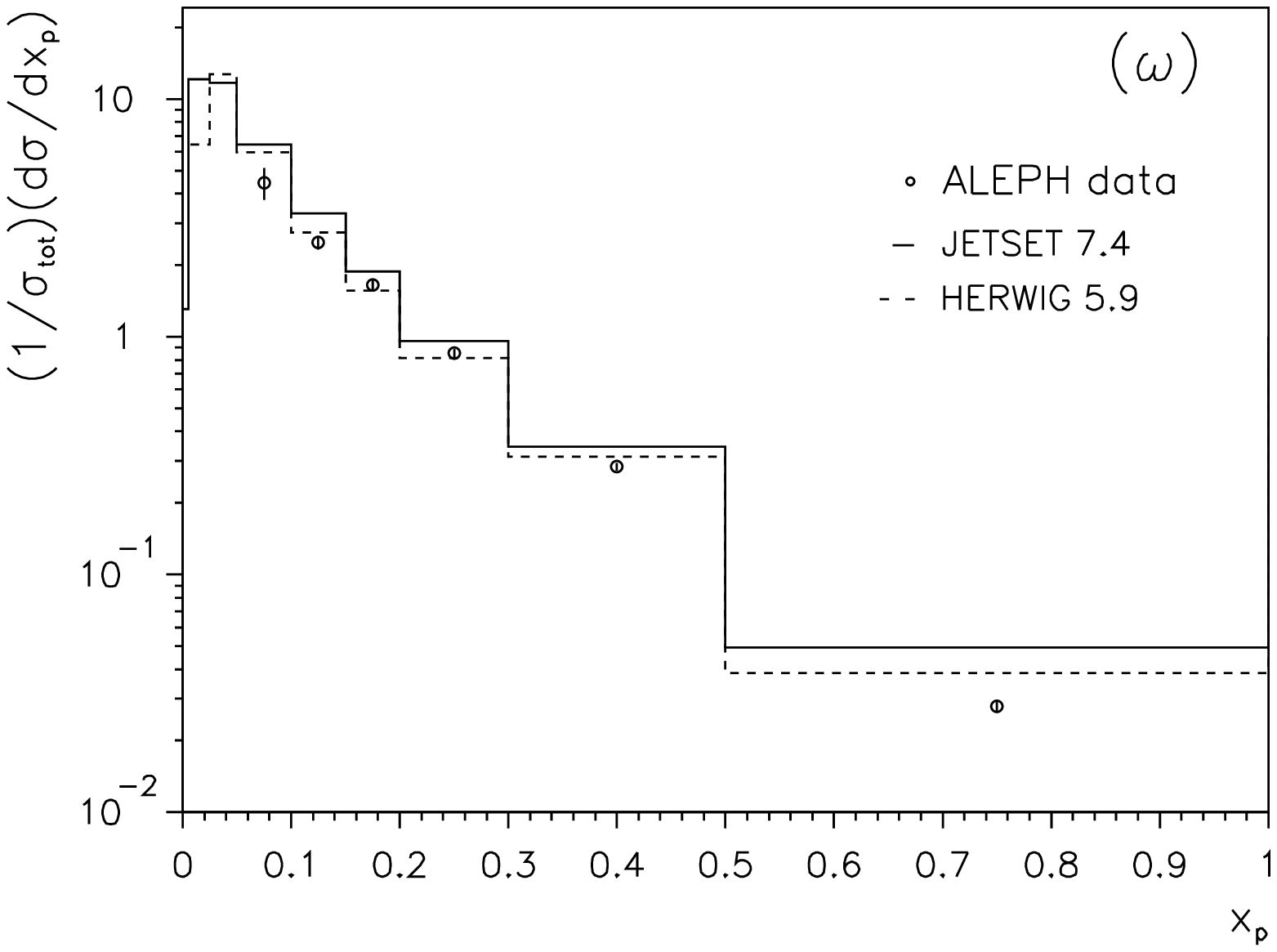,height=12cm}
\end{center}
\vspace{-50mm}
\caption[\small Measured $\omega$ cross section in comparison with 
        MC predictions.]
        {\label{OME_RESULTS}
        {\small Measured differential cross sections for the $\omega$
          in comparison with \MC predictions. The errors shown are the 
          quadratic sum of statistical and systematic contributions. }}
\end{figure}

Table~\ref{TAB_RESULTS} shows the results for
rates and differential cross sections in each measured momentum interval.
Results of summing over the measured $x_p$ intervals are shown; the
final line gives the result of extrapolating this to $x_p=0$
together with an extra error of 5.8\% and 2.4\% representing the uncertainty in the
extrapolation of the measured $\eta$ and $\omega$ rates, respectively.
Measured differential cross sections are compared to 
\MC predictions in Figs.~\ref{ETA_RESULTS} and \ref{OME_RESULTS}
for the $\eta$ and $\omega$, respectively.
Table~\ref{TAB_TOTRATES} compares the total rate in the data 
to the \MC predictions. The errors are the quadratic sum
of the statistical and systematic contributions. 
The final column gives
the result of extrapolating the measured rate to $x_p = 0$. 
Values inside the brackets indicate the discrepancy between
\MC and data in terms of the total experimental error.
Table~\ref{TAB_ETA_ALO} compares the results of the $\eta$
measurement with results published by ALEPH~\cite{NEWETA},
OPAL~\cite{OPAL} and L3~\cite{L3neweta}. 
For the $\omega$,
Table~\ref{TAB_ALO} compares the results with those of published 
results by ALEPH~\cite{NVMP}, OPAL~\cite{OPAL} and L3~\cite{L3}.

\begin{table}[phtb]
\vspace{-3mm}
\begin{center}
\caption[\small Measured rates and differential cross sections.]
     {\label{TAB_RESULTS}
     {\small Measured production rates per hadronic event 
      and differential cross sections for the 
      $\eta$ and $\omega$. The errors correspond to statistical and
      systematic uncertainties.
      The value for the total systematic error is taken from
      Table~\ref{TAB_ETA_ERRORS} and Table~\ref{TAB_OME_ERRORS} for the
      $\eta$ and $\omega$ respectively, where correlations between errors
      are taken into account.
      The results of summing over the measured $x_p$ intervals are 
      given, including the extrapolation to the full $x_p$ range with
      an additional error representing the uncertainty in the extrapolation.}}
\vspace{3mm}
\begin{tabular}{|c|l|c|}
\hline
$x_p$  & ~~~~~~~~~~~$\eta$ rate & $1/\sigma_{\mathrm {tot}} 
                          \cdot d\sigma/d x_p$ \\
\hline
\hline
 0.10-0.15 & 0.1236 $\pm$ 0.0098 $\pm$ 0.0182   & 2.471 $\pm$ 0.195 $\pm$ 0.363 \\ \hline
 0.15-0.20 & 0.0753 $\pm$ 0.0042 $\pm$ 0.0077   & 1.506 $\pm$ 0.084 $\pm$ 0.154 \\ \hline
 0.20-0.30 & 0.0802 $\pm$ 0.0028 $\pm$ 0.0047   & 0.802 $\pm$ 0.028 $\pm$ 0.047 \\ \hline
 0.30-0.50 & 0.0611 $\pm$ 0.0014 $\pm$ 0.0032   & 0.306 $\pm$ 0.007 $\pm$ 0.016 \\ \hline
 0.50-1.00 & 0.0146 $\pm$ 0.0004 $\pm$ 0.0009   & 0.029 $\pm$ 0.001 $\pm$ 0.002 \\
\hline \hline
 0.10-1.00 & \multicolumn{2}{l|} {0.355 ~$\pm$ 0.011 ~~$\pm$ 0.024 } \\ \hline
 all       & \multicolumn{2}{l|} {1.200 ~$\pm$ 0.037 ~~$\pm$ 0.082 $\pm$ 0.070} \\
\hline
\multicolumn{3}{l}{}               \\
\hline
$x_p$  & ~~~~~~~~~~~$\omega$ rate & $1/\sigma_{\mathrm {tot}} 
             \cdot d\sigma/d x_p$ \\
\hline
\hline
 0.05-0.10 & 0.2222 $\pm$ 0.0173 $\pm$ 0.0289   & 4.444 $\pm$ 0.347 $\pm$ 0.578 \\ \hline
 0.10-0.15 & 0.1246 $\pm$ 0.0052 $\pm$ 0.0065   & 2.493 $\pm$ 0.105 $\pm$ 0.130 \\ \hline
 0.15-0.20 & 0.0825 $\pm$ 0.0027 $\pm$ 0.0042   & 1.650 $\pm$ 0.054 $\pm$ 0.084 \\ \hline
 0.20-0.30 & 0.0852 $\pm$ 0.0024 $\pm$ 0.0033   & 0.852 $\pm$ 0.024 $\pm$ 0.033 \\ \hline
 0.30-0.50 & 0.0569 $\pm$ 0.0011 $\pm$ 0.0017   & 0.284 $\pm$ 0.006 $\pm$ 0.009 \\ \hline
 0.50-1.00 & 0.0139 $\pm$ 0.0004 $\pm$ 0.0008   & 0.028 $\pm$ 0.001 $\pm$ 0.002 \\
\hline \hline
 0.05-1.00 & \multicolumn{2}{l|} {0.585 ~$\pm$ 0.019 ~~$\pm$ 0.033             } \\ \hline
 all       & \multicolumn{2}{l|} {0.996 ~$\pm$ 0.032 ~~$\pm$ 0.057 $\pm$ 0.024} \\ 
\hline
\end{tabular}
\end{center}
\end{table}


\begin{table}[phtb]
\vspace{-0mm}
\begin{center}
\caption[\small Comparison of the total measured production rates with
         the \MC predictions.]{\label{TAB_TOTRATES}
        {\small Comparison of the total measured production rates 
        with \MC predictions.
             Values in parentheses indicate the discrepancy between
             \MC and data in terms of the total experimental error. }}
\vspace{3mm}
\begin{tabular}{|c|c|c|}
\hline
                  & ($\eta$) $x_p > 0.10$ & ($\eta$) all $x_p$  \\ \hline
\hline
            data  & 0.355 $\pm$ 0.026     & 1.20 $\pm$ 0.11     \\ \hline
 {\sc Jetset 7.4} & 0.296 $(-2.3 \sigma)$ & 1.00 $(-1.8 \sigma)$\\ \hline
 {\sc Herwig 5.9} & 0.289 $(-2.5 \sigma)$ & 1.04 $(-1.5 \sigma)$\\
\hline
\multicolumn{3}{l}{}               \\
\hline
              & ($\omega$) $x_p > 0.05$   & ($\omega$) all $x_p$           \\ \hline
\hline
        data  & 0.585 $\pm$ 0.038         & 0.996 $\pm$ 0.070     \\ \hline
 {\sc Jetset 7.4} & 0.771 $(+4.9 \sigma)$ & 1.310 $(+4.5 \sigma)$ \\ \hline
 {\sc Herwig 5.9} & 0.678 $(+2.4 \sigma)$ & 1.125 $(+1.8 \sigma)$ \\ \hline
\end{tabular}
\end{center}
\end{table}

%
\begin{table}[phtb]
\begin{center}
\caption[\small $\eta$ rate measured in this analysis compared to
         values published by LEP experiments.]{\label{TAB_ETA_ALO}
    {\small $\eta$ rate measured in this analysis compared to
         values published by LEP experiments.}}
\vspace{3mm}
\begin{tabular}{|r|l|l|}
\hline
              Experiment                   & $\eta$ decay mode                 & $\eta$ rate    \\ \hline 
                      L3   \cite{L3neweta} & $\gamma \gamma$                   & $0.93  \pm 0.01  \pm 0.09  \pm 0.06$ \\ \hline
                    OPAL   \cite{OPAL}     & $\gamma \gamma$, $\pip \pim \piz$ & $0.97  \pm 0.03  \pm 0.10  \pm 0.04$ \\ \hline
              this study                   & $\pip \pim \piz$                  & $1.20  \pm 0.04  \pm 0.08  \pm 0.07$ \\ \hline \hline
      ALEPH  \cite{NEWETA} $x_p > 0.10$   &  $\gamma \gamma$                   & $0.282 \pm 0.006 \pm 0.015         $ \\ \hline
 this study $x_p > 0.10$                   & $\pip \pim \piz$                  & $0.355 \pm 0.011 \pm 0.024         $ \\ \hline
\end{tabular}
\end{center}
\end{table}

\begin{table}[phtb]
\begin{center}
\caption[\small $\omega$ rates as measured by three LEP 
        experiments.]{\label{TAB_ALO}
        {\small Results for the $\omega$ production rate measured by 
             three LEP experiments. }}
\vspace{3mm}
\begin{tabular}{|r|l|}
\hline
 Experiment               & $\omega$ rate \\ \hline 
      ALEPH   \cite{NVMP} & $1.07 \pm 0.06 \pm 0.12 \pm 0.04$ \\ \hline
         L3   \cite{L3}   & $1.17 \pm 0.09 \pm 0.15         $ \\ \hline
       OPAL   \cite{OPAL} & $1.04 \pm 0.04 \pm 0.13 \pm 0.03$ \\ \hline
 this study               & $1.00 \pm 0.03 \pm 0.06 \pm 0.02$ \\ \hline
\end{tabular}
\end{center}
\end{table}

For the $\eta$, the total estimated production rate of
$1.200 \pm 0.037_{\mathrm {stat}} \pm 0.082_{\mathrm {sys}}
\pm 0.070_{\mathrm {extrap}}$ per event is somewhat higher than 
the predictions of both {\sc Jetset 7.4}
(1.00 $\eta$ per event) and {\sc Herwig 5.9} (1.04 $\eta$ per event).
This result is also somewhat higher than the results
published by LEP experiments.
The two measurements for the $\eta \rightarrow \gamma \gamma$
mode~\cite{NEWETA} and the $\eta \rightarrow \pip \pim \piz$
mode differ by more than twice the estimated error, 
the ratio (rate obtained with the $\pip \pim \piz$ mode to the
rate obtained with the $\gamma \gamma$ mode)
being $1.26 \pm 0.04_{\mathrm {stat}} \pm 0.10_{\mathrm {sys}}$.
This is consistent with a recent finding by OPAL \cite{OPAL}
of $1.14 \pm 0.07_{\mathrm {stat}} \pm 0.13_{\mathrm {sys}}$
for the same ratio.

For the $\omega$, the total estimated production rate of
$0.996 \pm 0.032_{\mathrm {stat}} \pm 0.057_{\mathrm {sys}}
\pm 0.024_{\mathrm {extrap}}$ per event lies significantly below
the prediction of {\sc Jetset 7.4} (1.31 $\omega$ per event) and
somewhat below that of
{\sc Herwig 5.9} (1.13 $\omega$ per event).
Published results from L3 and OPAL for the production rate of the
$\omega$ are in good agreement with the result of this analysis.


\section{\label{2MU}
    The Branching Ratio BR\boldmath $(\omega \rightarrow \mm)$}

\subsection{\label{2MU_EXTRACT}
            Extraction of the Signal}

\begin{figure}
\vspace{-10mm}
\begin{center}
\epsfig{file=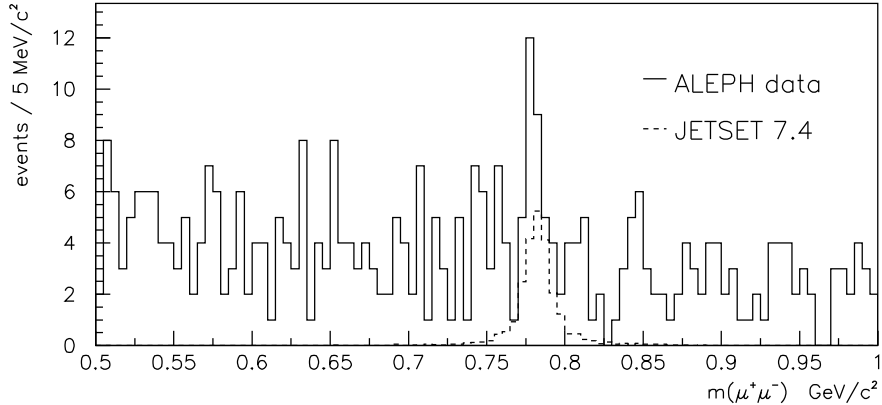,width=13.cm}
\end{center}
\vspace{-10mm}
\caption[\small Invariant mass distribution 
        $m(\mu^+ \mu^-)$]{\label{omumu-figsoft}
        {\small Invariant mass distribution $m(\mu^+ \mu^-)$ for hadronic 
         events containing a muon pair 
         after standard cuts as described in the text. 
         The $\omega$ signal
         of {\sc Jetset} is normalized to the signal in the data shown.}}
\end{figure}

The signal $\omega \rightarrow \mu^+ \mu^-$ is obtained from the
invariant mass distribution of two identified muons of opposite charge
originating from a common vertex (Fig.~\ref{omumu-figsoft}).
%
%
%
The signal is taken as the number of events in the mass region
$770\unit{MeV}/c^2 < m(\mu^+ \mu^-) < 795\unit{MeV}/c^2$
reduced by the estimated background in the same region.
Monte Carlo studies have shown the background to be linear as a function
of the invariant mass. The level of background in the signal region is
extrapolated from the background in the sidebands in the mass ranges 
$500\unit{MeV}/c^2 < m(\mu^+ \mu^-) < 745\unit{MeV}/c^2$ and
$820\unit{MeV}/c^2 < m(\mu^+ \mu^-) < 1000\unit{MeV}/c^2$.
Applying the cuts described in Sections \ref{EvSe} and \ref{PID} results 
in an optimal
signal-to-background ratio, $S/\sqrt{S+B}$. 
\begin{figure}
\vspace{-10mm}
\begin{center}
\epsfig{file=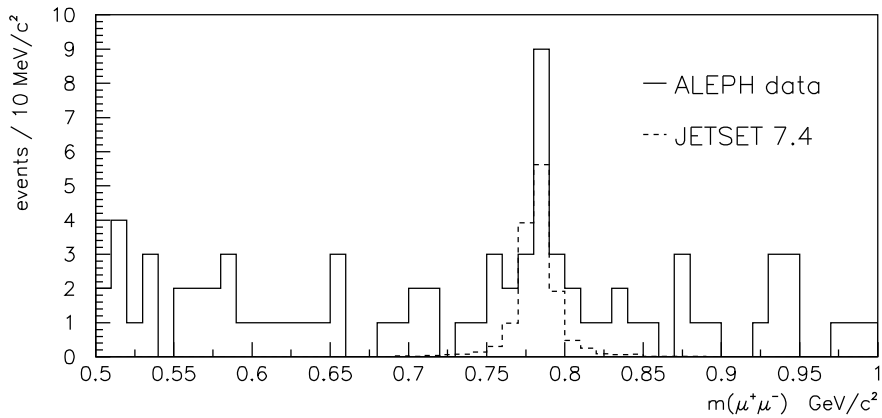,width=13.cm}
\end{center}
\vspace{-10mm}
\caption[\small Invariant mass distribution 
        $m(\mu^+ \mu^-)$]{\label{omumu-fighard}
        {\small Invariant mass distribution $m(\mu^+ \mu^-)$ for hadronic 
         events containing a muon pair 
         with tighter cuts (as explained in the text). The $\omega$ signal
         of {\sc Jetset} is normalized to the signal in the data shown.}}
\end{figure}

\subsection{\label{2MU_BR}
            Measurement of BR\boldmath $(\omega \rightarrow \mu^+ \mu^-)$}

Figure~\ref{omumu-figsoft} shows 35 events in the signal region. A 
background of 16.94 events is estimated from the sidebands, leaving 
$18.06 \pm 5.92$ signal events.
When the acceptance is taken from {\sc Jetset}~7.4, but with the momentum 
spectrum and rate corrected to agree with the $\pi^+\pi^-\pi^0$ analysis, 
a muonic branching ratio 
BR$(\omega \rightarrow \mu^+ \mu^-) = (9.0 \pm 2.9) \times 10^{-5}$ 
is obtained, based on the selected events shown in Fig.~\ref{omumu-figsoft}. 

The effect of tightening cuts in order to increase $S/\sqrt{B}$ is shown
in Fig.~\ref{omumu-fighard}. For this, the additional requirements are 
muon momenta greater than $2.5\unit{GeV}/c$, two VDET hits instead of one,
and a stronger $\bbbar$ event rejection (keeping 90\% of light 
quark-antiquark events).
In the signal region 14 events are kept, with an estimated 
background of 3.24 events, leaving $10.76 \pm 3.74$ signal events.

\subsection{\label{2MU_SYS}
            Systematic Error Analysis}

Uncertainties on the muonic branching ratio come from the uncertainty of the 
momentum spectrum of the $\omega$ and its rate in hadronic events, which are 
taken from the $\pi^+\pi^-\pi^0$ analysis. If the momentum spectrum 
from {\sc Jetset}~7.4 is used, the extracted branching ratio 
would increase. Half of the difference from the shape of the spectrum 
is taken as systematic error (4\%). 
Further systematic errors come from the uncertainty of the measured rate of
$\omega$ production per event in the $\omega \rightarrow \pi^+\pi^-\pi^0$
analysis. Because event selection and track selection are similar in
both analyses their errors are assumed
fully correlated and neglected. The remaining
errors ($\pi^0$ selection, fitting procedure, etc.) are studied by varying the
Monte Carlo momentum spectrum according to these remaining errors.
An error of 6\% is obtained.

Possible uncertainties can arise from the track selection and b-tag 
probability. The last also changes the flavour composition of the 
sample; the $\omega$ rate for different flavours has not yet been
measured. The $\pi^+\pi^-\pi^0$ analysis was therefore repeated with
the cuts on b-tag probability and track selection  from this analysis. 
An error of 4\% is derived. 
The estimation of the background is dominated by the limited 
statistics in the sidebands. Smaller uncertainties come from a possible 
curvature of the background shape, e.g., 
due to the $\rho$ meson. 
The error is determined to be 6\%. Further systematic uncertainties 
are related to the performance 
of the ALEPH detector \cite{PERFORMANCE}: muon identification 
\cite{muonid} ($\pm 5\%$), 
tracking resolution, which influences the signal shape ($\pm 5\%$), and 
the missing energy, the impact of which is found to be negligible. 
The limited Monte Carlo statistics adds 3\% to the error.
Other sources of systematic error such as 
background are found to be small and are neglected.
The total systematic error is obtained by adding the contributions in
quadrature, giving $\pm 13\%$.

\subsection{Results}

The muonic branching ratio is measured to be 
BR$(\omega \rightarrow \mu^+ \mu^-) = (9.0 \pm 2.9 \pm 1.1) \times 10^{-5}$ 
from the observation of $18.06 \pm 5.92$ signal events. 
This is in agreement with the expectation from the flavour composition 
of the vector mesons and from lepton universality.

\section{Summary and Conclusion}

The inclusive production of the $\eta$ meson and the 
$\omega$ vector meson in  hadronic Z decays
has been studied and compared to model predictions. 
Decays of the type $\eta \rightarrow \pip \pim \piz$
are reconstructed for the momentum interval of $x_p > 0.10$
where $x_p = p_{\mathrm {meson}}/p_{\mathrm {beam}}$.
Decays of the type $\omega \rightarrow \pip \pim \piz$
are reconstructed for the momentum interval of $x_p > 0.05$. 
A signal is seen in the muonic decay $\omega \rightarrow \mu^+ \mu^-$. 

The average $\eta$ rate per event for $x_{p} > 0.10$ is measured to be
$0.355 \pm 0.011_{\mathrm {stat}} \pm 0.024_{\mathrm {sys}}$;
an extrapolation to $x_p = 0$ using the shape of the fragmentation
function in {\sc Jetset 7.4} yields a total production
rate of $1.200 \pm 0.037_{\mathrm {stat}} \pm 0.082_{\mathrm {sys}} \pm 
0.070_{\mathrm {extrap}}$ per event. 
The predictions of both {\sc Jetset 7.4} (1.00 $\eta$ per event)
and {\sc Herwig 5.9} (1.04 $\eta$ per event)
are somewhat below this result.
The use of the matrix elements proposed by \cite{OPAL} and
\cite{carpenterME} resulted in a significant effect on the
momentum spectrum for low $\eta$ momenta.
The matrix element is not present in {\sc Jetset 7.4} and {\sc Herwig 5.9}.
Published results of L3 \cite{L3neweta} and OPAL \cite{OPAL} 
are somewhat lower than the result of this analysis. 

The average $\omega$ rate per event for $x_{p} > 0.05$ is measured to be
$0.585 \pm 0.019_{\mathrm {stat}} \pm 0.033_{\mathrm {sys}}$;
an extrapolation to $x_p = 0$ yields a total production
rate of $0.996 \pm 0.032_{\mathrm {stat}} \pm 0.057_{\mathrm {sys}} 
\pm 0.024_{\mathrm {extrap}}$
per event. The rate lies significantly below the prediction of
{\sc Jetset 7.4} (1.31 $\omega$ per event) and somewhat below that of 
{\sc Herwig 5.9} (1.13 $\omega$ per event).
Published results from L3 and OPAL for the production rate of the
$\omega$ are in good agreement with the result of this analysis.
This measurement also agrees with that previously published in
\cite{NVMP} and improves it substantially.

In the muonic decay mode of the $\omega$ vector meson, $18.1 \pm 5.9$
events are observed yielding the first measurement of the muonic 
branching ratio of BR$(\omega \rightarrow \mu^+ \mu^-) = 
(9.0 \pm 2.9_{\mathrm {stat}} \pm 1.1_{\mathrm {sys}}) \times 10^{-5}$.
The result is in agreement with the expectation from the flavour composition 
of the vector mesons and from lepton universality.

\subsection*{Acknowledgements}

We wish to thank our colleagues in the CERN accelerator divisions for
the successful operation of LEP. We are indebted to the engineers and
technicians in all our institutions for their contribution to the
excellent performance of ALEPH. Those of us from non-member
countries thank CERN for its hospitality.


\end{document}